\documentclass[longbibliography,showpacs,floatfix,nofootinbib,superscriptaddress,twocolumn,aps,prb]{revtex4-2}

\usepackage{graphicx,color}
\usepackage{amsfonts}
\usepackage[figuresright]{rotating}  
\usepackage{amssymb}
\usepackage{amsmath}
\usepackage{mathtools}
\usepackage{psfrag}
\usepackage{tabularx}
\usepackage{textcomp}
\usepackage{units}
\usepackage{graphicx}% for "\scalebox" macro
\usepackage{physics}
\usepackage{times}
\usepackage[colorlinks=true,citecolor=blue,linkcolor=magenta,hypertexnames=false]{hyperref}
\usepackage{enumitem}
\usepackage[caption=false]{subfig}
\usepackage{tikz}
\usepackage{appendix}
\usepackage{dsfont}

\renewcommand{\selectlanguage}[1]{}

\newcommand{\sx}{\sigma^x }
\newcommand{\sy}{\sigma^y }

\newcommand{\sigp}{\sigma^+ }
\newcommand{\sigm}{\sigma^- }

\makeatletter
\newcommand*{\transpose}{%
	{\mathpalette\@transpose{}}%
}
\newcommand*{\@transpose}[2]{%
	\raisebox{\depth}{$\m@th#1\intercal$}%
}

\newcommand{\ra}{\rangle}

\renewcommand{\ket}[1]{| #1 \ra}

\newcommand{\LL}{\mathcal{L}}

\newcommand{\AAA}{\mathcal{A}}
\newcommand{\DD}{\mathcal{D}}

\newcommand{\NN}{\mathcal{N}}

\renewcommand{\dd}{\text{d}}
\newcommand{\ii}{\text{i}}
\newcommand{\ee}{\text{e}}

\newcommand{\numC}{\mathbb{C}}

\renewcommand{\unit}{\mathds{1}}

\begin{document}

\title{On the non-integrability of driven-dissipative one-dimensional hard-core bosons}

\author{M. Zündel}
\email{martina.zuendel@lpmmc.cnrs.fr}
\affiliation{Univ. Grenoble Alpes, CNRS, LPMMC, 38000 Grenoble, France}
\affiliation{Kirchhoff-Institut für Physik, Ruprecht-Karls-Universität Heidelberg, INF 227, 69120 Heidelberg, Germany}

\begin{abstract}
We address the question whether hard-core bosons, equivalent to the XX-model, remain integrable once the system is no longer closed. 
We consider the lattice version under incoherent local pump and loss and show, using random matrix theory, that the statistics of the complex spacial ratios indicate that the system is chaotic. Further, we show that the model belongs in the AI$^{\dagger}$ universality class of random matrices. In addition to this analysis, we investigate an emergent stripe pattern in the Lindbladian spectrum and relate it to the dissipative parameters of the model.
\end{abstract}
\maketitle

\section{Introduction}
The Lieb–Liniger model~\cite{lieb_exact_1963} is a paradigmatic example of an integrable many-body system in one dimension~\cite{kinoshita_quantum_2006-1}. In the limit of infinitely strong interactions, it reduces to the Tonks–Girardeau gas, which enjoys an exact solution obtained through the Bose-Fermi mapping~\cite{girardeau_relationship_1960}. This even holds at finite temperatures and under real-time evolution~\cite{lenard_onedimensional_1966, girardeau_dynamics_2003,yukalov_fermi-bose_2005}. A natural follow-up question arises when the system is rendered {\em open}—that is, coupled to an external environment or reservoirs. Does such coupling preserve the integrable structure, or does it introduce effective interactions and decoherence that spoil integrability?

Open quantum systems can be realized in various ways~\cite{sieberer_universality_2023, gardiner_quantum_2004, breuer_open_2006, rivas_open_2012-1, vega_dynamics_2017}, such as through coupling to a thermal bath, measurement-induced decoherence, or engineered dissipation channels. One rigorous method for describing the dynamics of the open subsystem is the Lindblad -- Gorini-Kossakowski-Sudarshan master equation~\cite{gorini_completely_1976, lindblad_generators_1976}: This approach ensures that the time evolution is completely positive and trace-preserving~\cite{lindblad_generators_1976}, meaning that the density matrix remains physically valid at all times. As a result, the Lindblad framework is widely used to model dissipative processes in quantum mechanics, providing a consistent and microscopic description of Markovian-open system dynamics~\cite{rivas_open_2012-1}.

A straightforward numerical method to test whether a closed quantum system is integrable or chaotic is to examine the spectrum of the relevant time-evolution operator, such as the Hamiltonian.
Random matrix theory (RMT) predicts that the energy levels follow a Poissonian distribution for uncorrelated level spacings, indicating that the system is integrable,
while chaotic closed systems exhibit level repulsion characterized by Wigner–Dyson statistics. A more convenient diagnostic is the ratio of energy level spacings in closed systems have been introduced in~\cite{vadim_localization_2007,atas_distribution_2013}.
Furthermore, it has been conjectured~\cite{bohigas_characterization_1984} that for chaotic systems, the spectral statistics of the Hamiltonian are determined by its symmetry class.

An analogous approach to assess the integrability of open quantum systems involves analyzing the complex level spacings of the non-Hermitian operators governing their dynamics. A suggestion of analyzing the distributions of complex spacing ratios (CSR) has been put forward in~\cite{sa_complex_2020, sa_symmetry_2023}.
Their analysis suggests that if the distribution is Poisson-like, meaning flat in the complex plane, then the system is integrable - given all symmetries of the system were taken into account. Whereas, if the distribution follows a Ginibre ensemble, i.e., given the polar coordinates of the CSR, if one can observe cubic level repulsion in the marginal distribution of the radial coordinate for small radii and a non-isotropic angular distribution, then the system is not integrable. Since no exact formula exists for the surmise of the Ginibre ensembles in the large-$N$ limit, the comparison to a two-dimensional tori unitary ensemble (TUE) has been put forward~\cite{sa_complex_2020}.
Recently, an approximate formula for the CSR of the Ginibre class was provided~\cite{dusa_approximation_2022}.

However, the classification of non-Hermitian systems~\cite{bernard_classification_2002,kawabata_symmetry_2019, hamazaki_universality_2020, sa_symmetry_2023, kawabata_symmetry_2023, sa_symmetry_2024} goes beyond this. It has been shown~\cite{hamazaki_universality_2020} that non-Hermitian systems follow one of three distinct universality classes of random matrices AI, AI$^{\dagger}$, AII$^{\dagger}$.

In this work, we aim to infer the integrability of strongly interacting bosons in a Markovian environment using the concept of CSR.
To enable numerical investigation, we study systems with finite-dimensional Hilbert spaces. The Tonks–Girardeau gas is a continuum version of hard-core bosons (HCB) on a one-dimensional lattice, where the constraint that no more than one boson can occupy a site effectively mimics infinite repulsion. In contrast to the Bose-Hubbard model at finite interaction in one-dimension, the HCB model at infinite interaction is integrable; see for recent progress in calculating the spectral function see~\cite{rigol_emergence_2011,settino_exact_2021-1,patu_exact_2022-1}. Through the Holstein–Primakoff transformation~\cite{holstein_field_1940}, we map HCB to the XX spin-$\frac{1}{2}$-chain. This spin model is integrable and can be mapped with the Jordan-Wigner transformation~\cite{jordan_uber_1928} to free Fermions. Here, the Pauli exclusion principle takes care of the hard-core constraint. The equivalence of the Jordan-Wigner transformation and Girardeau’s Bose-Fermi mapping can be shown explicitly on a lattice~\cite{cazalilla_one_2011}.

\begin{figure*}[ht!]
     \centering
     \includegraphics[width=1.\linewidth]{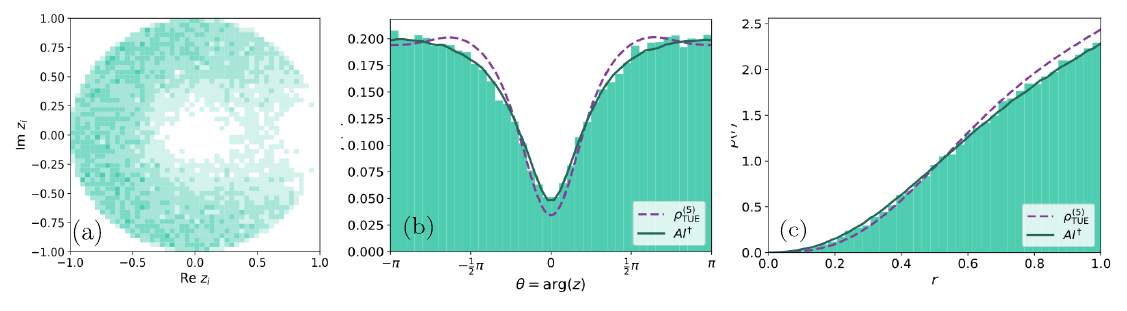}
     \caption{Analysis of the CSR~\eqref{eq:CSR} of the spectrum of the driven-dissipative XX spin-$\frac{1}{2}$ model~\eqref{eq:model}: (a) Distribution $\rho(z)$, (b) marginal distribution $\rho(\theta)$ over the angle $\theta$, (c) marginal distribution $\rho(r)$ over the radius $r$. In violet (dashed), the surmise for the TUE with $N=5$ is shown, this is an approximation for the GinUE, compare with App.~\ref{app:tue}. In dark green (solid), the sampled surmise for the symmetry class AI$^\dagger$ is displayed. Data of $104468$ eigenvalues, for $L=12$, $M=1$, at $\gamma_l=J=1, \gamma_p=0.8$ in the $0$-momentum and positive parity sector.}
     \label{fig1}
\end{figure*}

The paper is organized as follows: In section~\ref{sec:model}, we define the model of hard-core bosons on a lattice with local Markovian pump and loss and discuss its symmetries. In section~\ref{sec:csr}, the main results of this paper are shown: We present and analyze the CSR of an irreducible block of the Lindbladian, indicating that the model is not integrable as it does not follow a Poisson distribution, as investigated in~\cite{sa_complex_2020}. We classify the marginal distributions over angle and radius by comparison to RMT results. Section~\ref{sec:spectral} is independent of the discussion on integrability: We present features of the spectrum and compare them with perturbative calculations in the dissipative parameters.

\section{Model}\label{sec:model}
We consider a periodic chain of hard-core bosons
\begin{align}
    H_{HCB} = -J\sum_{j=1}^L (b_j^{\dagger}b_{j+1} + h.c.)\;,
\end{align}
with the hard-core constraint given by the two-dimensional local Hilbert space, such that $b_i^2\ket{\Psi}=0=b_i^{\dagger 2}\ket{\Psi}\,\forall\Psi$, where $b_i$ and $b^{\dagger}_i$ are the bosonic annihilation and creation operators, and $n_i = b^{\dagger}_ib_i$, the local boson number density. This can be mapped using the Holstein–Primakoff transformation~\cite{holstein_field_1940}
\begin{align}
    \sigp_j = b^{\dagger}_j \sqrt{1-n_j}\;,\quad
    \sigm_j = \sqrt{1-n_j} b_j\;,
\end{align}
to the XX spin-$\frac{1}{2}$-chain model given by the Hamiltonian
\begin{align}
    H_{XX} = -2J \sum_{j=1}^L (\sx_j \sx_{j+1} + \sy_j \sy_{j+1})\;,
\end{align}
with the physical spin $S^{\mu}=\frac{1}{2}\sigma^{\mu}$, where $\{\sigma^{\mu}\}_{\mu = x,y,z}$ are the Pauli matrices. We define $\sigma^{\pm} = \frac{1}{2} (\sx \pm \ii \sy)$, which act as spin-flip operators and it follows that the operator $n_{S,i} = \sigp_i\sigm_i$ returns the local number of excited states. The pump and loss of hard-core bosons hence correspond to spin flips in the XX-model. 

We choose to model the open system with the Lindblad master equation and argue further that the simplest model for homogeneous loss and gain over the system contains local, single operator terms. Whereas, higher order operators necessarily are non-local: Due to the hard-core nature, at most one particle can be placed or taken out in one position. Therefore, we choose the resulting Lindbladian to be of the form
\begin{align}\label{eq:model}
    \LL \rho = - \ii [H_{XX}, \rho] + \gamma_p \sum_{j=1}^L \DD[\sigp_j]\rho + \gamma_l \sum_{j=1}^L \DD[\sigm_j]\rho\;.
\end{align}
The dissipator $\DD$ is defined below and describes the spin-flip up (down) - pump (loss) of a boson - at rate $\gamma_p$ ($\gamma_l$).
Note that applying the Jordan–Wigner transformation to this Lindbladian renders it highly non-linear and non-local in the fermionic degrees of freedom, due to the Jordan-Wigner strings. Therefore, the model is not quadratic and cannot be solved as in~\cite{prosen_third_2008}.

\subsection{Vectorization}
In order to implement the time evolution operator $\LL$ as a matrix operator and extract its eigenvalues, we vectorize the operators as $A \rho B \mapsto (A \otimes B^{\transpose}) \rho$. The von Neumann equation reads
\begin{align}
    \partial_t \rho &= - \ii [H, \rho]\\
    &\mapsto - \ii \Big(
    H \otimes \unit - \unit \otimes H^{\transpose}
    \Big) \rho\;,
\end{align}
and the dissipator $\DD$ of the jump operator $L$
\begin{align}
    \DD[L] \rho &= L \rho L^{\dagger} - \frac{1}{2} L^{\dagger}L \rho - \frac{1}{2} \rho L^{\dagger} L\\
    &\mapsto \Big(
    L \otimes L^* - \frac{1}{2} L^{\dagger}L \otimes \unit -  \frac{1}{2} \unit \otimes L^{\transpose} L^*
    \Big) \rho\;,
\end{align}
with the asterisk $*$ denoting complex conjugation.
If the Lindbladian possesses symmetries or additional structure, we need to first block-diagonalize it and find spectra {\em per} sector. Only in a symmetry-reduced basis can we distinguish an integrable from a chaotic system.

\subsection{Symmetries}
In open systems, there are two types of possible symmetries -- weak and strong symmetries~\cite{buca_note_2012}.\footnote{Similar considerations have been made within the Keldysh formalism~\cite{sieberer_universality_2023}, calling them respectively {\em classical} and {\em quantum} symmetries.} A symmetry is called {\em strong} if the Hamiltonian $H$ and each jump operator $L_i$ separately commute with the associated operator $A$ of the symmetry
\begin{align}
    [H,A] = 0 \quad \land\; [L_i, A] = 0\;, \forall i\,.
\end{align}
A symmetry is called {\em weak} if 
\begin{align}
    \LL\big(A\rho A^{\dagger}\big)=A(\LL\rho)A^{\dagger}\;.
\end{align}
Strong symmetry implies weak symmetry.
Finally, we can {\em block-diagonalize} $\LL$ if a superoperator $\AAA \equiv [A, .] = A \otimes \unit - \unit \otimes A^{\transpose}$ commutes with the Lindbladian
\begin{align}
    [\LL,\AAA] = 0\;.
\end{align}

It can be easily shown that the model in Eq.~\eqref{eq:model} does not allow for strong conservation of the particle number (total magnetization in the spin picture). Nevertheless, the entire Lindbladian commutes with the corresponding super operator $[\LL,\NN]=0$, which, in the language of the hard-core bosons, is a super-particle number operator $\NN=n\otimes \unit - \unit \otimes n^{\transpose}$, with $n=\sum_i n_i$. Its eigenvalues $M\in\{-L,\dots,L\}$ depend on the difference of particles $M=N_m-N_n$ in the bra $\bra{N_m}$ and ket state $\ket{N_n}$ describing the density matrix $\Phi_{m,n}=\ket{N_m}\bra{N_n}$.

Further, the momentum operator is conserved in a strong way, manifesting that the system is translational-invariant. Additionally, the system displays a discrete strong spatial parity symmetry  $b^{(\dagger)}_i\mapsto b^{(\dagger)}_{L-i+1}$ ($\sigma^{\mu}_i\mapsto\sigma^{\mu}_{L-i+1}$), when in the momentum sector $0$ or $\pi$.
Finally, we note that in the particular case of $\gamma_p=\gamma_l$, the particle number sector $M=0$ has additional symmetry between the empty and occupied site (a spin flip symmetry). 

\section{Analysis of the level ratio}\label{sec:csr}
The CSR $z_i$ for the complex eigenvalue $\lambda_i\in\numC$ is defined as the ratio of the distances to the nearest $\lambda^{(NN)}_i$ and next-nearest eigenvalue $\lambda^{(NNN)}_i$,
\begin{align}\label{eq:CSR}
    z_i = \frac{\lambda_i-\lambda^{(NN)}_i}{\lambda_i-\lambda^{(NNN)}_i}\;.
\end{align}
Representing the ratio $z$ in polar coordinates $z=r\ee^{\ii \theta}$, we can analyze the marginal distributions over the angle $\rho(\theta)=\int_0^1 \dd r\, r \rho(r,\theta)$ and radius $\rho(r)=\int_0^{2\pi}\dd\theta\, r \rho(r,\theta)$.

For non-Hermitian systems, the CSR $z$ for chaotic systems are expected to follow the Ginibre surmise~\cite{hamazaki_universality_2020}. Interestingly, for small radii, all three Ginibre ensembles (GinUE, \mbox{GinOE}, GinSE), show cubic level repulsion, hence they are indistinguishable by what is known as Dyson index $\beta$ in closed systems.

Since the low-$N$ Ginibre ensembles do not describe the large-$N$ limit well, we make use of two-dimensional tori ensemble introduced in~\cite{sa_complex_2020}. The idea is to equally distribute the eigenvalues over a torus, parametrized by the angles $\vartheta\in(-\pi,\pi],\varphi\in(-\pi,\pi]$,
\begin{align}
    &P^{(N)}_{TUE}(\vartheta_1, \dots, \vartheta_N; \varphi_1, \dots, \varphi_N)\\
    &\quad\propto 
    \prod_{j<k} \Big(2-\cos(\vartheta_j-\vartheta_k)-\cos(\varphi_j-\varphi_k)\Big)\;,
\end{align}
in order to take into account the strong finite size boundary effects. The surmise of the TUE, the analog of the Wigner surmise, can be found in App.~\ref{app:tue} and is converges fast for low $N$.

The implementation is explained in App.~\ref{app:numerics}. We fixed the super-particle number sector, as well as the momentum and parity sector. The results are displayed in Fig.~\ref{fig1}(a). The CSR over the complex plane shows clearly an inhomogeneous distribution, with a hole in the middle, as compared to a flat Poisson distribution, hence hinting at chaotic behavior of the system. The asymmetry in the angle distribution, representing a level repulsion of the eigenvalues, underlines this further. For more quantitative results the marginal distributions over the angle and radius are shown in Fig.~\ref{fig1}(b),(c), and compared to the afore mentioned tori ensemble in violet. The qualitative comparison is matching -- hence, we conclude that the system is not integrable.

In order to understand the slight quantitative mismatch, we take into account the different symmetry classes the generator of the open system dynamics can fall into~\cite{hamazaki_universality_2020}. Following their nomenclature, we see that the surmise of the symmetry class AII$^{\dagger}$ would increase the discrepancy to the model data further. This can be seen for instance in the lowered minimum of $\rho(\theta)$. However, the opposite behavior is found the class AI$^{\dagger}$.

Therefore, we additionally compare our data to the surmise of the symmetry class AI$^{\dagger}$. For this symmetry class there exists no analytical expression for the distribution or surmise, contrary to the class AI. For this reason, we sample $2^{12}\times2^{12}$ random matrices under the constraint of their class, calculate their CSR and average over $2^{10}$ realizations, to obtain the surmise marked in dark green. The surmise well describes the CSR of this symmetry sector of the model. 

\section{Properties of the spectrum}\label{sec:spectral}
In this section, we further investigate the Lindbladian spectrum of the model Eq.~\eqref{eq:model}.
\subsection{Numerical observations}
 \begin{figure}[h!]
     \centering
     \includegraphics[width=1.\linewidth]{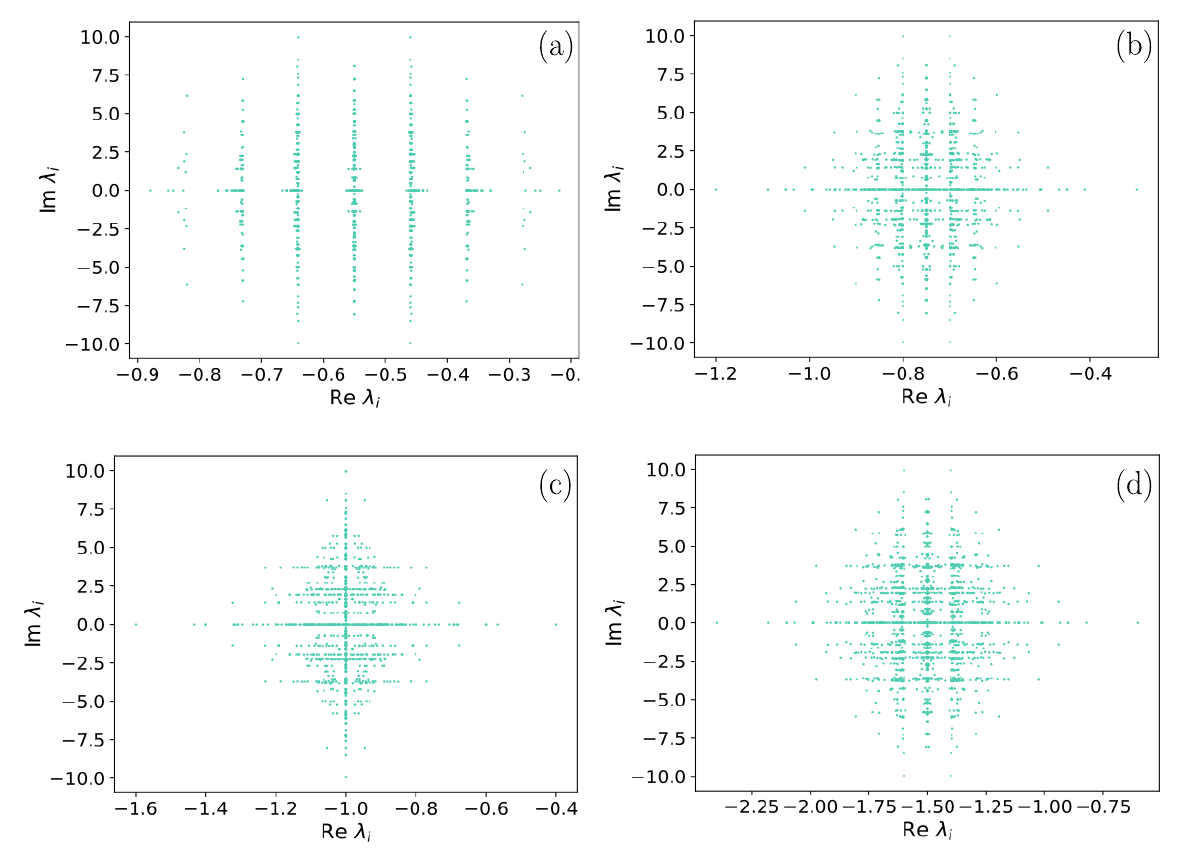}
     \caption{Spectrum of the Lindbladian, for system size $L=10$, in the super particle number (magnetization) sector $M=4$, with varying pump rate $\gamma_p=0.01,\,0.05,\,0.1,\,0.2$ from (a)-(d), $\gamma_l=0.1$ and $J=1$.}
     \label{fig2}
 \end{figure}
\subsubsection*{Change of the pump rate}
We find that varying the pump rate $\gamma_p$ separates different parts of the spectrum, compare Fig.~\ref{fig2}. Setting the pump rate equal to the loss rate yields the spectrum shown in Fig.~\ref{fig2} (c). The lines of the real axis and of the imaginary axis, shifted by the trace of the Lindbladian, are still pronounced. For smaller and larger pumping rates, segments of the spectrum begin to separate, leading to an apparent stripe-like pattern. The stripes are separated by a uniform distance.
We observe this stripe pattern even for pump and loss rates of order $1$ or higher. The following perturbative prediction, cf. Sec \ref{sec:pt1}, does not hold quantitatively in this regime, but captures nonetheless the emergent stripe pattern.

\subsubsection*{Apparent stripes}
\begin{figure}[h!]
     \centering
     \includegraphics[width=1.\linewidth]{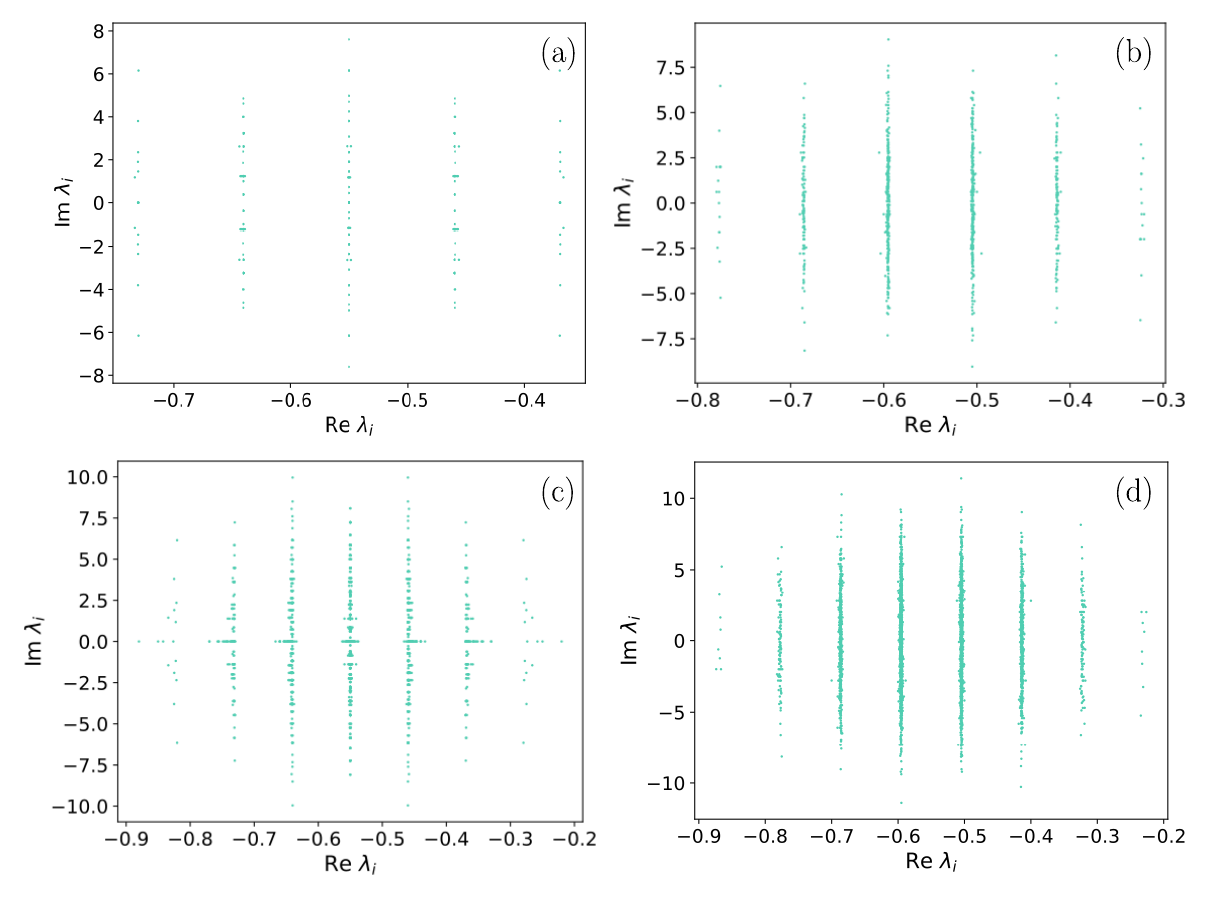}
     \caption{Spectrum of the Lindbladian, for system size $L=10$, in different super particle number (magnetization) sectors $M=6,\,5,\,4,\,3$ from (a)-(d). Model parameters $\gamma_p=0.01$, remaining parameters as in Fig.~\ref{fig2}.}
     \label{fig3}
 \end{figure}

Fig.~\ref{fig3} shows that the number of visible stripes depends on the chosen super-particle number sector $M$ used to diagonalize the matrix. This number appears to be given by $L-|M|+1$. We cross-checked this with other system sizes $L$. Naturally, for fewer eigenvalues, the stripe shape eventually vanishes. However, the number of clusters (stripes) stays the same. For instance, when $L=10$ and $M=9$, only two states exist whose separation can be tuned with the pump rate $\gamma_p$, and $M=10$ with exactly one state in it.

\subsection{Perturbative prediction}\label{sec:pt1}
 
 \begin{figure}[h!]
     \centering
     \includegraphics[width=.9\linewidth]{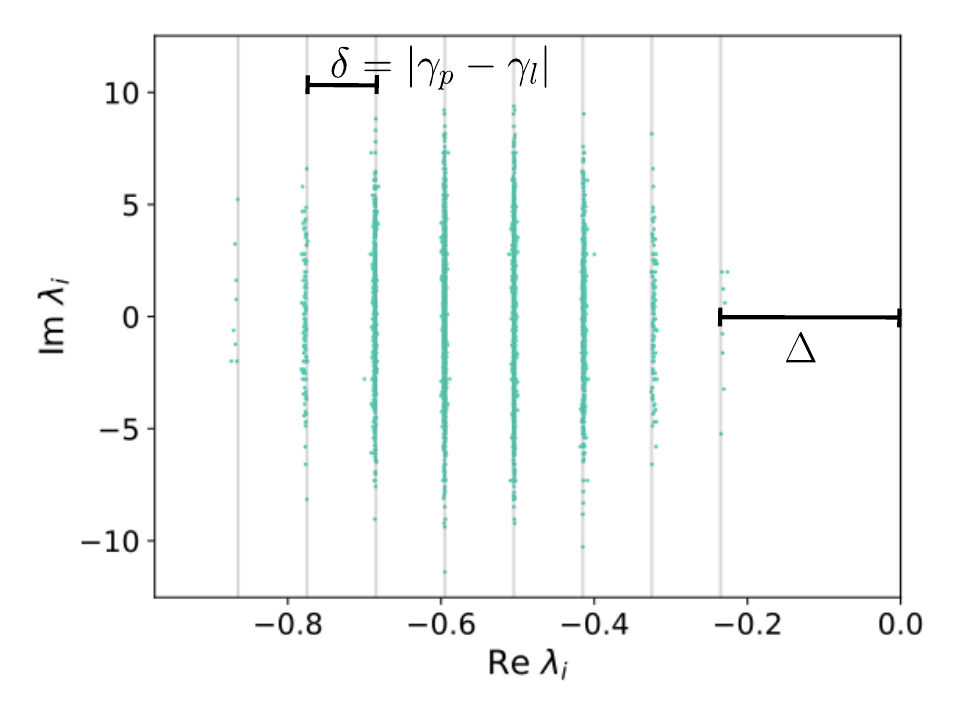}
     \caption{Spectrum of the Lindbladian for system size $L=10$ and super particle number (magnetization) sector $M=3$. Model parameters: $J=1$, $\gamma_p=0.01$, $\gamma_l=0.1$. Perturbative prediction~\ref{sec:pt1} of the position of the stripes in gray. To first order in perturbation theory: Separation of the stripes $\delta=|\gamma_p-\gamma_l|$, distance to the origin $\Delta=-\gamma_pL+\frac{1}{2}(\gamma_p-\gamma_l)M$.}
     \label{fig3b}
 \end{figure}

Assuming that the dissipative parameters are small $\gamma_p,\gamma_l\ll1$, we can split the Lindblad super-operator
\begin{align}
    \LL = \LL_0 + \gamma \LL_1\;,
\end{align}
into an unperturbed part $\LL_0$, consisting of the coherent, von Neumann evolution
\begin{align}
    \LL_0 \rho = -\ii[H_{XX}, \rho]\;,
\end{align}
and treat the dissipative part as a small perturbation
\begin{align}
    \gamma \LL_1 = \gamma_p \sum_{j=1}^L \DD[\sigp_j]\rho + \gamma_l \sum_{j=1}^L \DD[\sigm_j]\rho\;.
\end{align}
We will calculate the first-order corrections $\lambda^{(1)}_{mn}$ to the unperturbed eigenvalues $\lambda^{(0)}_{mn}$.

Let $\ket{E_m}=\ket{\{m_j\}_{j=1,\dots,L}}$ be the Fock basis, which is an eigenbasis of the Hamiltonian $H_{XX}$ with $H_{\mathrm{XX}} \ket{E_m} = E_m \ket{E_m}$. We call the right eigenoperator of the unperturbed Lindbladian $\LL_0$
\begin{align}
    \Phi_{m,n} &= \ket{E_m}\bra{E_n}\;,
\end{align}
and find its unperturbed eigenvalues
\begin{align}
  \lambda_{m,n}^{(0)}
  &=
  -\ii \,(E_m - E_n)\;,
\end{align}
are purely imaginary as expected. The first-order shift in that eigenvalue, for a non-degenerate case, is given by
\begin{align}\label{eq:ev1_formula}
  \lambda_{m,n}^{(1)}
  &=
  \frac{
    \langle\!\langle \tilde{\Phi}_{m,n}, \,\mathcal{L}_1(\Phi_{m,n}) \rangle\!\rangle
  }{
    \langle\!\langle \tilde{\Phi}_{m,n},\,\Phi_{m,n} \rangle\!\rangle
  }\;,
\end{align}
where $\langle\!\langle A,B\rangle\!\rangle = \Tr(A^\dagger B)$ is the Hilbert-Schmidt inner product in operator space, and $\tilde{\Phi}_{m,n}$ is the left eigenoperator of $\LL_0$ corresponding to $\Phi_{m,n}$. The eigenoperators are normed $\langle\!\langle \tilde{\Phi}_{m,n},\,\Phi_{m,n} \rangle\!\rangle =1$.

Evaluating the numerator of Eq.~\eqref{eq:ev1_formula}, we see that only the second and third term in the dissipator $\tfrac12 \{L^{\dagger}L,\rho\}$ give a contribution to the diagonal elements of the operator and hence to the first-order correction in the eigenvalues.
Summing over the lattice sites, we denote $N_m = \sum_i \bra{E_m} n_i \ket{E_m}$ as the total number of excitations in state $\ket{E_m}$. Similarly $N_n$ for $\ket{E_n}$. Then the first-order correction to the eigenvalues is given by
\begin{align}
  \lambda_{m,n}^{(1)}
  &= -\gamma_pL
  +\tfrac12(\gamma_p-\gamma_l)(N_m + N_n).
\end{align}
This corresponds to a negative shift of the imaginary unperturbed eigenvalues $\lambda^{(0)}_{m,n}$ along the real axis. This shift depends on the number of excited states in the bra $N_m$ and ket $N_n$ of $\Phi_{m,n}$.

We show the theoretical prediction together with the data of the model for small pump and loss rate in Fig.~\ref{fig3b}. Due to the fixed super particle number sector $M=N_m-N_n= \text{const}$, the smallest difference $\delta$ of two distinct shifts is given by $\delta=|\gamma_p-\gamma_l|$, explaining the regularly spaced stripes, observed in the numerical calculations and described in the previous section.

The number of stripes is equally determined by the particle number sector $M$ within a system of size $L$. There can be $0$ particles in $\ket{E_m}$, follows $M$ particles in $\ket{E_n}$, or $1$ and $M+1$, etc., up to $L-M$ and $L$ particles. In total $L-|M|+1$ possibilities, which gives the number of stripes conjectured in the previous section. We conclude that the decay, that is the negative real part of the eigenvalues, is extensive in the number of excitations $N_n+N_m$.

Calculating higher order corrections in the dissipative parameters could further unravel the exact shape of the distributions centered at the line positions predicted from the first order result. For instance, to second order, off-diagonal elements with the weight $\gamma_p\gamma_l$ enter the calculation. The second order can be associated to the width of the stripes. Comparing this to the separation of the stripes $\gamma_p\gamma_l/(\gamma_l-\gamma_p)$, it is small for the loss and pump rate being very different, $\gamma_p\ll\gamma_p$ or visa versa, explaining the appearance of the stripes.

\section{Conclusions}\label{sec:conclusion}
In this paper we showed that a system of hard-core bosons on the lattice loses its integrable structure if losses and gains of the system are implemented with bulk Lindbladian one-body jump operators. We did so by considering the CSRs of the spectrum of the generator of the dynamics. These showed clearly a donut-shape in the complex plane and anisotropic angular distribution, as expected for a chaotic system~\cite{sa_complex_2020}. Furthermore, we find that the chosen irreducible representation of the system to be in the universality class AI$^{\dagger}$~\cite{kawabata_symmetry_2019} of random matrices. 
In a separate part, we analyzed emergent features of the Lindbladian spectrum of this model. At large ratio of pump ans loss rate, we found a number of $L-|M|+1$ regularly spaced stripes. For small dissipative parameters, a perturbative calculation confirmed and explained these patterns. We determined the Lindbladian gap as well as the line positions that correspond to different average relaxation time scales. For a purely dissipative, random Liouvillian with some degree of locality, similar clusters have been found~\cite{piazza_2020}.

In summary, we provide a clear answer to the question of integrability of a driven-dissipative system of strongly interacting lattice-bosons coupled to an environment by local loss and gains. Further, the result suggests the continuum version, the Tonks-Girardeau gas, to be non-integrable; As the Jordan-Wigner transformation, responsible for the mapping to free fermions in the closed system, fails to maintain the integrability once the system is open, the same is to be expected for the Bose-Fermi mapping in the continuum limit.

We stress that this seems to us the most straightforward and experimentally relevant way of implementing an {\em open} bosonic system within the Lindbladian framework. Such a bosonic lattice with local pump and loss can be realized in exciton‐polariton condensates in semiconductor microcavities~\cite{Kasprzak2006}, in optomechanical systems with local feedback control~\cite{ludwig_quantum_2013}, in superconducting circuit QED arrays of qubits and resonators~\cite{raftery_observation_2014,gourgy_quantum_2016}, or in ultracold‐atom optical lattices with site‐selective addressing~\cite{barotini_controlling_2013}.
We note, however, that other prominent examples with non-local two-body losses exist, specifically nearest neighbor hopping terms, among which there exist integrable models.
For instance, in~\cite{de_leeuw_constructing_2021}, Yang-Baxter integrable open systems are constructed. Within this construction, they find the known integrable case of the XX model with dephasing noise~\cite{medvedyeva_exact_2016}, as well as the totally antisymmetric and symmetric simple exclusion process, and discuss the generalizations thereof. A recent publication~\cite {sa_symmetry_2024} puts this in context with the non-Hermitian symmetry classifications. Excluding bulk drive and loss, another integrable model involving two boundary-connected replicas has recently been found~\cite{paletta_2025}.

The question of the existence of further integrable and experimentally relevant solutions for open bosonic systems remains open.

\textit{Acknowledgments}--- The author warmly thanks Anna Minguzzi for discussions on related topics, Tomaž Prosen for helpful advice, and Lucas Sà for correspondence. Further, the author thanks Léonie Canet and Sebastian Diehl for their encouragement to publish the results and Anna Minguzzi, Anton Khvalyuk, Félix Helluin, and Sébastien Lucas for proofreading. This work is supported in parts by the German Research Foundation (Deutsche Forschungsgemeinschaft, DFG) under Germany's Excellence Strategy EXC 2181/1- 390900948 (the Heidelberg STRUCTURES Excellence Cluster).

\appendix

\section{Details on the analytical distributions}\label{app:tue}
\begin{figure}[h!]
     \centering
     \includegraphics[width=1.\linewidth]{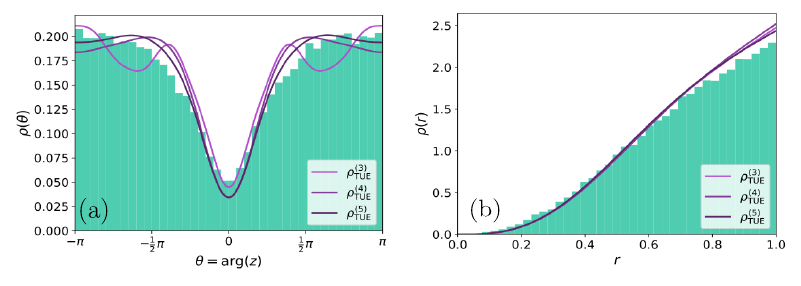}
     \caption{Marginal distributions $\rho(\theta)$ in (a), $\rho(r)$ in (b), of the TUE's surmise at $N=3,4,5$. Together with the data of Fig.~\ref{fig1}.}
     \label{fig4}
 \end{figure}

The surmise of the TUE is given by~\cite{sa_thesis}:
\onecolumngrid
\begin{align*}
\rho_{TUE}^{(N)}(x,y) 
&\propto
\int_{-\pi}^{\pi}\!\! \int_{-\pi}^{\pi} ds\,dt 
\;\prod_{j=1}^{N-3} \bigl(\int_{-\pi}^{\pi}da_j\,\int_{-\pi}^{\pi}db_j\bigr)\;
\Theta\bigl((a_j^2 + b_j^2)\;-\;(s^2 + t^2)\bigr)\,
(s^2 + t^2)^2\,[2 - \cos s \;-\; \cos t]\\
&\quad\times\;
[2 \;-\; \cos(sx - ty) \;-\; \cos(tx + sy)]
\,[2 \;-\; \cos\bigl(s(x-1) \;-\; ty\bigr) \;-\; \cos\bigl(t(x-1) + sy\bigr)]\\
&\quad\times\;
\prod_{j=1}^{N-3}
\bigl[\,2 \;-\; \cos a_j \;-\; \cos b_j\bigr]\;
\bigl[\,2 \;-\; \cos(s - a_j) \;-\; \cos(t - b_j)\bigr]\\
&\quad\times\;
\bigl[\,2 \;-\; \cos(sx - ty - a_j) \;-\; \cos(tx + sy - b_j)\bigr]
\;\prod_{j<k}\;
\bigl[\,2 \;-\; \cos(a_j - a_k) \;-\; \cos(b_j - b_k)\bigr].
\end{align*}
\twocolumngrid
We evaluate the integrals numerically with Monte Carlo sampling and show the marginal distributions $\rho(r)$ and $\rho(\theta)$ of the TUE's surmise for $N=3,4,5$ in Fig.~\ref{fig4}.

\section{Construction of the Lindbladian in the symmetry reduced basis}\label{app:numerics}
The Lindbladian matrix was constructed in a symmetry-reduced basis, analogous to the Hamiltonian in closed systems. Because the super-particle number $\NN$ commutes with the generator of the dynamics $\LL$, first, all basis elements of a fixed sector $M$ were selected. Further, we constrain the matrix to the basis of semi-momentum states as we choose the zeroth momentum sector $P=0$ and the positive parity sector. Compare~\cite{sandvik_computational_2010} for an implementation of semi-momentum states in closed system dynamics. 

We chose the system size reasonably large $L=12$ and picked the sector $M=1$ to obtain good statistics with $104468$ eigenvalues.

We performed two more checks: 
First, performing the calculation in the momentum sector $P=0$, but not in the semi-momentum states, leads to a spectrum resembling Poisson statistics. This indicates integrability or the disregard of a symmetry, as expected since the parity symmetry was not taken into account.
Second, we choose the sector $M=0$ (in a slightly smaller system $L=11$), which leads to an additional spin flip parity symmetry. This sector is of particular interest because it contains the steady state of the dynamics. Again, as expected, not correctly taking the symmetry into account leads to something rather close to the Poisson distribution.

\bibliographystyle{apsrev4-2}

\bibliography{library}

\clearpage

\end{document}